\documentclass[12pt]{article}

\author{ Vladimir A. Petrov }

\title{The idea of quarks: towards restoring of historical justice.}
\date{}

\begin{document}

\maketitle
A. A. Logunov Institute for High Energy Physics, NRC KI,
Protvino, RF

\begin{abstract}
Quite a long time ago several authors (see,e.g., \cite{Der}, \cite{Pe}  ) mentioned that among pioneers of the quark idea we should take into consideration one more (in addition to M. Gell-Mann and G. Zweig) author, Andr\'{e} Petermann (1922 - 2011). Below we place the English translation of his little known work, originally published in French.
We draw attention of the readers to the closesness of the dates: M. Gell-Mann's paper in Phys.Letters was received 4 January 1964, the CERN preprint by G. Zweig is dated by 17 
January 1964 while Petermann's paper in  Nucl.Phys. was received  30 December 1963.
\end{abstract}
\begin{flushright}
\textbf{{\Large {Nuclear Physics 63 (1965) 349-352}}}
\end{flushright}

\begin{center}
PROPERTIES OF STRANGENESS  AND A MASS FORMULA FOR VECTOR MESONS
\end{center}

\begin{center}
A.Petermann
\end{center}
\begin{center}
CERN, Genève 23
\end{center}

\begin{center}
Received 30 December 1963
\end{center}

\textbf{Abstract}:\textit{ A mass formula for vector mesons is proposed, and the role of strangeness in mass formulae is discussed.}

   In spite of success won by the formula of Gell-Mann–Okubo \cite{Ge}  giving the mass difference of ordinary baryons (nucleons, $\Lambda, \Sigma, \Xi  $) it proved incapable to give clarification on the mass differences of vector mesons. The aim of the present note is to compensate this gap and give, at least on a trial basis, a dynamic explanation of the role which plays the strangeness in this mass formula.                                                                              
   
   Let us consider to this end two spinors $ s $ and $ s^{'} $ (their anti-spinors are designated by $ \bar{s} $ and $ \bar{s}^{'} $), being $ s^{'} $ is strange ( $ \mid S\mid =1 $) while $ s $ is not. Electromagnetic and weak interactions are decoupled in order to consider only the world of interactions stronger than they. It should be noted that if the electromagnetic interaction would be present we should be obliged to consider 3 spinors   $ s,\hat{s} $ and  $ \hat{s}^{'} $ , i.e. the isospinor $ (\hat{s}, s) $ with S = 0 and the isoscalar $ \hat{s}^{'} $ with $ \mid S\mid =1 $. The simplified case results in the degeneracy $ s = \hat{s} $ which manifests itself in the absence of the electromagnetic field.
It is known that strangeness is conserved similarly to charge: we call it the “strange charge” expressed in units that we will specify below. This charge is carried without loss by $ s^{'} $, which ensures its conservation. Such a conservation law may be derived from a principle of local gauge invariance($ \psi_{s^{'}} \rightarrow \psi _{s^{'}} e^{-iS\lambda(x)} $) . To ensure the invariance of the Lagrangian of $ \psi_{s^{'}} $ under such a gauge transformation it is necessary to introduce a neutral vector field $ \Sigma_{\mu} (x) $ transforming adequately when changing gauge. This field is naturally coupled only to the field $ \psi_{s^{'}} $. Dynamically, it will produce a self-energy effect with a self-mass proportional  to  $ S^{2}(\sigma^{2}/4\pi) $. Experimentally, the mass differences between particles  of  different strangeness, such as those between $ \Lambda $ and $ N $ or $ \Xi $ and $ \Sigma $, show us that they can be regarded as a perturbation in comparison with the absolute value of the mass of these particles. Recalling then the effect of the field $ \Sigma_{\mu} $ , an elementary calculation similar to the purely electromagnetic case shows us that self-masses by relative order of magnitude are consistent with those from the perturbation series in  $ \sigma^{2}/4\pi $. 
 As the first order of self-mass effect is positive, it follows that the particle $ \bar{s}^{'} $ has a mass greater than that of $ s $ , so one can say that
 
\textbf{an elementary spinor particle of strangeness $ (\sigma^{2}/4\pi)^{1/2} $   is heavier than that same spinor particle without strangeness.}

In addition, if $ S^{2}(\sigma^{2}/4\pi) $ were the real coupling constant, then there would be a law of increase in mass proportionally  to the square of strangeness, while experimentally the increase is linear:
\begin{center}
$1/2(\Xi + N) \approx (1/2)(\Sigma + \Lambda).  $
\end{center}
This obliges us to assign the values $ S=0 $ and $\mid S\mid = 1 $ to the strangeness of $ s $ and $ s^{'} $ , respectively ($ 0 $ and $ 1 $ for $ \bar{s} $ and $ \bar{s}^{'} $  ), and to consider the particles of higher strangeness ($ \Xi $,  for example) as composed of particles $ s s^{'} $, but not as elementary. Then, holding the principle that, when neglecting, in addition to electromagnetism and weak interactions, the  strangeness, ordinary baryons offer a complete degeneracy ( $ N=\Lambda = \Sigma =\Xi) $); if $ \Xi $  is a particle composed of $ s $ and $ s^{'} $  then other baryons are too. In order to build them by means of $ s $ and $ s^{'} $  no  $ \bar{s} $ and $ \bar{s}^{'} $ can take part in this construction because, otherwise, the pair $ \hat{s}^{'} (s^{'}) $  could intervene, so that ordinary baryons of strangeness  $ S = 0 $ could be heavier than  ordinary baryons of higher strangeness, which is excluded experimentally\footnote{The presence of  $ \bar{s} $  is also excluded if one wants to build all the families of known baryonic   particles and nothing more.} . So that they must be composed of at least three $ s $ without $ \bar{s} $ , since there is a baryon of strangeness  $|S|= 2$. Thus we find the linear increase in mass with strangeness. (The comparison may be useful with effects that, in the nuclei, are proportional to $ Ze^2 $  rather than to $ Z^2 e^2 $ , explained by the fact that the nuclei are conglomerates of $ Z $  charged particles and not of particles of charge $ Ze $ ).We then see that the so-called elementary particles such as $ N,\Xi, \Sigma $ etc., are complicated objects, actually the states of strongly bound elementary spinor particles. But dynamics is beyond the scope of the current theory of fields.
We do not insist on these ordinary baryons because they provided the input to the model above described in the following circumstances we recall:

(1) determined in the order of $ (\sigma^{2}/4\pi)^{1/2} $  ;

(2) the observed law of linear increase of their masses with strangeness.

On the other hand, according to its title, this note should focus on vector mesons. These are, naturally, composed in this model of one $ s $  and one $ \bar{s} $ . Vector properties of $ s $  (a trivector  with components, $ s, \hat{s} $ and  $ s^{'} $  if the electromagnetic interaction is not decoupled) make it possible, by combination  $ s \bar{s}$, the formation of 9 vector mesons in accordance with observation.
If electromagnetism is neglected, their composition in terms of $ s $ can be specified symbolically as follows:

\[\rho= s\bar{s}, \:K^{\ast} =s^{'}\bar{s},\: \varphi = s^{'}\bar{s}^{'},\: \omega = s\bar{s}.\]

From the identity
\[2m_{s^{'}} \equiv 2m_{s^{'}} + 2m_{s} - 2m_{s}\]

we obtain the mass formula
\begin{equation}
  m_{\varphi} = 2m_{K^{\ast}} - m_{\rho}
 \end{equation}                                          
which could be remarkably tested in experiment if to take into account the experimental  errors in values of these masses and show also that the particle of strangeness $ 0 $ ,$ \varphi $, has larger mass than the particle with strangeness $ |S|= 1 $ and one could say:

\textbf{while for ordinary baryons mass increases with $ |S| $, this is not the case for vector mesons.} 
This fact was one of the major shortcomings in applying the Gell-Mann-Okubo formula.
 Furthermore, if we introduce a parameter $ Z $, like, for example, the number of
protons in a nucleus and which here specifies the number of $ s{'} $ and $ \bar{s}{'} $ participating in the composition of a vector meson, a mass formula for these can be written in the form 

\[m_V=A+BZ,\]

$ m_V $ being the mass of the vector meson.

Constant A can me immediately taken from experiment : this is the mass of the $ \rho $ –meson for which $Z=0  $. So, $ A= m_\rho $ . Constant B is the difference of masses between  particle $ s^{'} $  and    particle $ s $ . From the baryonic mass spectrum, namely from that of the baryonic resonances, 
the examination of which in the light of present consideration is under investigation , B may be estimated as $ 140 \pm 20 $ MeV. 
This estimate ignores, as was done during of the establishment of Eq. (1), the difference in binding energy between $ s^{'} $  and $ \bar{s}^{'}$, on the one hand, and $ s $ and $ \bar{s} $ on the other, in front of the absolute value of the binding energy and masses involved. One then obtains

\begin{flushleft}
$(1)\; Z=0: m_{\rho} = m_{\varphi},$
\end{flushleft}
\begin{flushleft}
$(2)\; Z=1: m_{K^{\ast}} = m_{\rho} + (140 \pm 20) MeV,$
\end{flushleft}
\begin{flushleft}
$(3)\; Z=2: m_{\varphi} = m_{\rho} + 2(140 \pm 20) MeV = m_{K^{\ast}}+ (140 \pm 20) MeV.  $
\end{flushleft}

 The agreement with the observed values is good if we take into account the experimental errors that are most likely greater than the change affecting the binding energies between $ s^{'}\bar{s}^{'} $  and  $ s\bar{s} $  and neglected as it has been said above.
In conclusion, one can make the following comments:

(1) The model is by the force of things rather crude since it cannot account for the difference in binding energies evoked.

(2) When the electromagnetic interaction is present, difficulties with
the electrical charge are present, either in the form of  non-conservation of the latter when the particle  $ s $ is binded to form particles that can be observed in the physical world. 
Or, if one wants to preserve the conservation of charge, which is highly desirable, particles $ s $ should then have non-integer values of the charge. This fact is unpleasant but cannot, after all, be excluded on physical grounds. Other means are, perhaps, possible to overcome this difficulty.
Contrariwise, the assumption of strange field as a corollary of conservation
of  strangeness provides a dynamic explanation for a strange particle
heavier than a particle without strangeness, whatever the sign of strangeness of that particle. Therefore, the model carries a rigorous logic. It provides particularly accurate information on mass differences of vector mesons and their curious dependence on strangeness.

(\textit{Translated from French by V.A. Petrov.})

\end{document}